# Transformation from Identity Stone Age to Digital Identity

Mohit Kohli,

Security Practices-Rolta Technology Limited (RIL), India

*Mohit.kohli@gmail.com, Mohit.kohli@rolta.com*

## ABSTRACT

*Technological conversion, political interests and Business drivers has triggered a means, to establish individual characterization and personalization.*

*People started raising concerns on multiple identities managed across various zones and hence various solutions were designed. Technological advancement has brought various issues and concerns around Identity assurance, privacy and policy enabled common Authentication framework. A compressive framework is needed to established common identity model to address national needs like standards, regulation and laws, minimum risk, interoperability and to provide user with a consistent context or user experience.*

*This document focuses on Transformation path of identity stone age to Identity as in state. It defines a digital identity zone model (DIZM) to showcase the Global Identity defined across the ecosystem. Also, provide insight of emerging Technology trend to enable Identity assurance, privacy and policy enabled common Authentication framework.*

## 1. INTRODUCTION

Identity is a means to represent authenticity of an object. For humans, identity can be shown via birth certificate, passport, SSN etc. Similarly, Objects/Assets can be validated through serial numbers or bar-codes. We are only going to explore Human Identities and their relationships.

In early days, Information technology was not mature to support digital forms of identity and associated relationships. Technological conversion, Political interests and Business drivers have triggered other means, to establish individual characterization and personalization.

Identity representation has evolved over the generations and can be categorized as Identity Stone Age, Interim Phase of Digital Identity and Future Digital Identity

**Identities Stone Age** refers to era where person needs to physically prove his existence either via close associate or references. In early decades, human used to prove existence using reference, via token object or paper based proof. The major concern during this phase was management of proof, huge resource utilization for validations, secure storage and corruption. There were no disaster management practices in place, due to the physical nature of the proofs. Identity revolution has transformed the meaning and has gone through many phases.

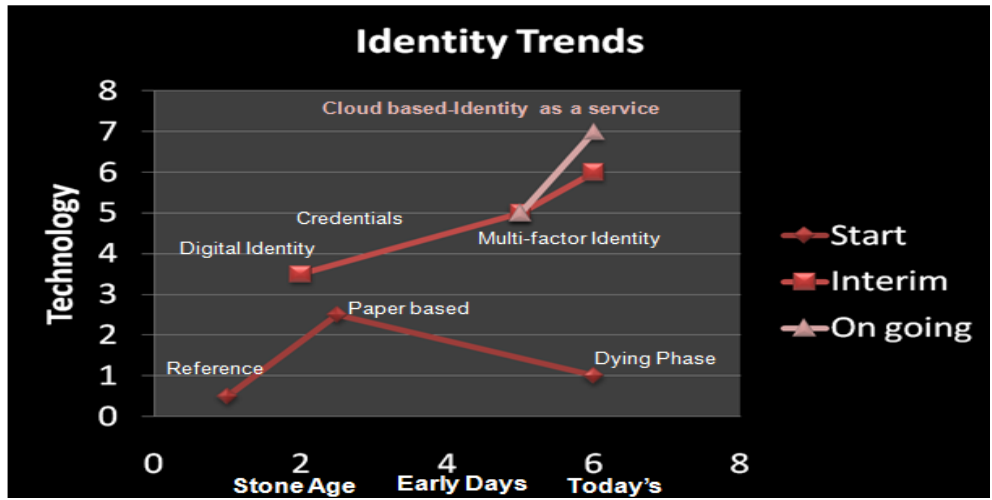

Figure 1 : Identity Trend

Digital Identity is a means to electronically represent or prove authenticity of individual. This is an integral part of digital technologies (Applications, systems, device etc.) and provides relevance to prove real people over the internet. However; there are some short falls like Internet identity misuse, Fake proofing etc.

Today, Identity can be either of the following digital forms-Credentials, RFID (RFID in the Aerospace and Defense Market), Strong Authentication-Token based, Token less, Smart card, OTP, Portable Id-Card space, Open Id(Used for online authentication), Claim id, naymz, PKI etc. Cloud computing, Identity as a service (Idas) and Authentication as a service (AAS) are some of the emerging terminologies, which are going through the initial research and legalization phase.

**Identity Attributes**-Credentials OR Biological Attributes OR Algorithm based OR Paper

This equation highlights the different means user can choose to access the information. The transition from one state to another is a gradual process which took many decades.

In this paper, we are going to address various Business & user requirement around sharing of information in an ecosystem with higher reliability and confidentiality. An identity & role equation will derive in context with digital identity Zone model. Various solutions will be analyzed when its compared with global identity needs of users in line with universal registration process, authentication mechanism, user ease and sharing of authorization data between service provider's and Identity provider's.

## 2. RELATED WORKS

With the superfluous use of internet and its application, it's becoming intensively difficult to identify the real user. And, user has to hunt for different credentials to access applications and data segregated across. Digital identity hacking [15][21], has raised various concerns and survey suggest around 9 million cases of identity theft were reported in the United States alone.

- 1 in Every 10 American consumer has been a Victim of Identity Theft
- 1.6 Million Households have had their bank accounts and/or debit cards Compromised
- The Average Amount Taken from each Identity Theft Victim to $4,841
- Nearly 50% OF victims learn of their identities being stolen within 3 months
- 25.0 million Americans now carry identity theft insurance

As a result a comprehensive Identity management systems was developed, aim to facilitate the task of identity management and to ease the control of identity information for authorized entities, as well as helping to preserve user privacy. The last few years have witnessed a significant growth in the number of identity management systems, and this number is expected to grow further in the next few years. An RNCOS2 report [23] predicts that the identity management market will grow at a compound annual growth rate of nearly 23% between 2009 and 2012. The vast majority of these identity management systems are not interoperable, and implementation and privacy issues remain. This thesis aims to enhance the privacy and practicality of identity management systems.

The delegation service framework for the Liberty Alliance project takes advantage of the trust relationships that exist by definition within the Liberty Alliance circles of trust, and involves the use of delegation assertions that can be built using the Security Assertion Mark-up Language (SAML) version 2 standard. SAML based solution [9] was introduce to provide secure authentication of user between 2 entities. Phillip J. Windley [19], represented the definition of digital identity, protocols for creating, exchanging, and using digital identity, and to develop an identity management strategy in your business.However, the solution has scalability concerns and not a feasible solution for internet ecosystem having vast range of application.

The consumer registration patterns [24] and behavior suggest around 75% are bothered by website registrations process and will change their Behavior as a result. 45% admit to leaving a website instead of re-setting their password or answering security questions. As a result this analysis will provide a mechanism to build user friendly common registration process.

OpenID [4] was created in the summer of 2005 by an open source community trying to solve a problem that was not easily solved by other existing identity technologies. As such, OpenID is decentralized and not owned by anyone, nor should it be. OpenID is rapidly gaining adoption on the web, with over one billion OpenID enabled user accounts and over 50,000 websites accepting OpenID for logins from many major Providers [2].

The JISC Final Report [7], allow decision-makers to understand OpenID's security properties in order to perform risk assessment of their envisaged use cases and avoid any of OpenID's potential security pitfalls. The project conducted a survey of computer centre managers and senior staff members to gain an understanding of how they are likely to proceed with OpenID, with or without the presence of this guidance. The secondary aim was to develop bridging software, the OpenID-SAML Gateway, to allow OpenIDs from any source to be used as identities within the production UK federation, creating opportunities for experimentation by early adopters.

There are many such mechanisms [19] available in the market to manage user identity in global market.

## 3. TODAY'S WORLD

By the time Nations started driving towards Digital identity many thefts mechanisms were awaken to steal identities. Modern technology solution was struggling to prevent identity thefts to enable secure identity infrastructure. Service Organizations started anticipating the need of Identity management solution to secure the online transactions and services.

In the nurture phase of digital identities, organizations were also facing major concerns around identity silos distributed across environments. The segregated Role and privilege association model has raised angst around Manageability, Tacking and costing.

By now, many forms of **Independent** identity management solutions were introduced to manage *Identity laws and establish interaction between Roles and services.*

### 3.1 Digital Identity zone model (DIZM)

The *Digital Identity zone model (DIZM),* describe various zones where identity is managed, for example Friends Family Zone-People are eager to share information with their "friends" in social networks like Orkut, Facebook, in chat rooms, or in Second Life), Purchase Zone – Customer are taking advantage of various offerings using EBay/Telecom services, Cooperate-Employees performing task on application infrastructure, Service-Banks started expending online transactions through Account banking, loan etc(Banking can be a classic example where entity may have multiple identities which are associated) .We are going to use the same model for rest of the analysis.

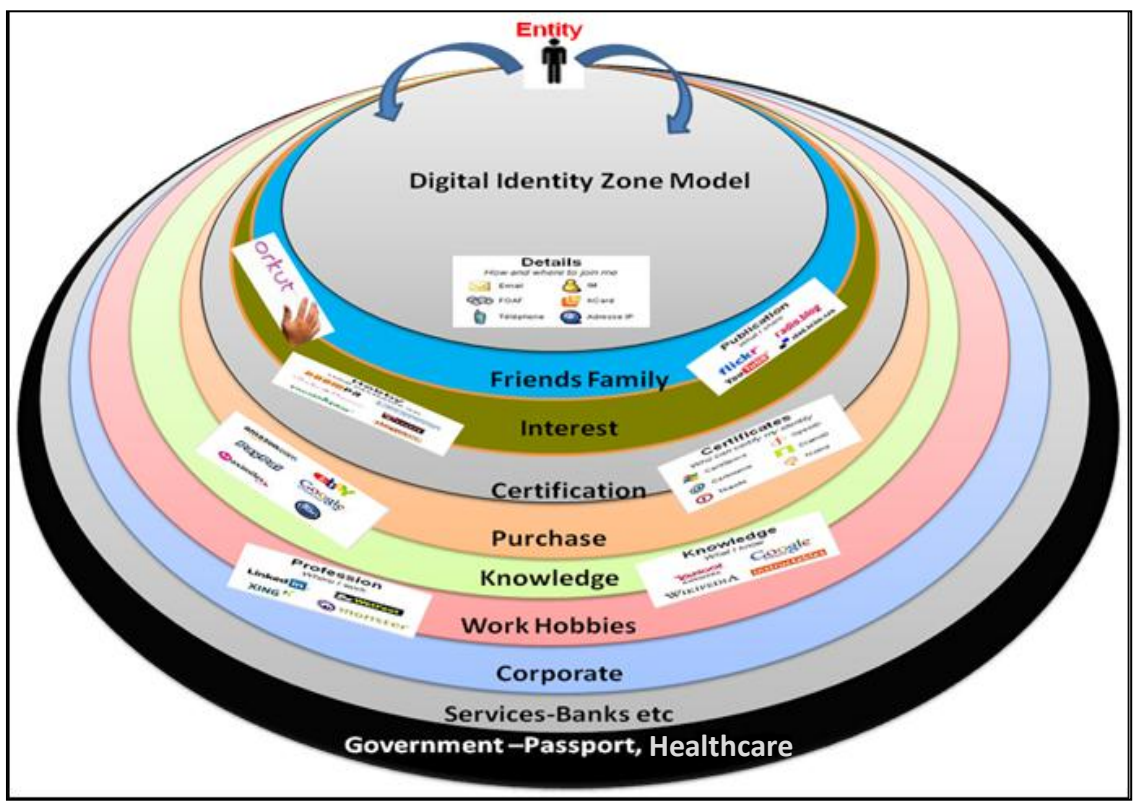

Figure 2: Digital Identity Zone Model

### Identity Equation in today's world would be:-

User = Many Identities =Many Roles=Many Resources=Many Access Mechanism

**Information Data Classification-**

1. Confidential- Government, Purchase, Healthcare
2. Private-Corporate
3. Sensitive-Services-Banks, Certification
4. Public-Work Hobbies, knowledge, interest, Friends Family.

"Identity in various zones may or may not be synchronized to support consolidated identity prototype". Identity synchronization can be cluster based on the zone, for example, all identities dealings with government agencies), others we may deliberately want to keep separate (such as identities used for online banking and access to our medical records) etc. Identity across the entire zone provides the monolithic architecture model; where universal identity can be establish to carry the interaction.

Now, organization started analyzing Identity model which can define interaction between various Identity sub sets .A general model for identity can be constructed from the trends highlighted below.

### 3.2 Trends

- ✓ An entity may have many Identities with Associated roles, services; which may be federated to share security access privileges. The conceptual relationship can be established from the below Paradigm

- ✓ In a real world, Telecom is also one of the zones (Purchase for Customers), which required integration of Identity with the sets of desired services. Abstract identities are defined uniquely based on exclusivity of identities and its sub sets. The services are mapped to the users and associated roles in order to provide access control based on the credential profiles. Telecom customer has multiple identities (Account Id's) which are mapped to the service roles. For Example Account holders, delegated account holders etc. Delegated account holder role is used to delegate services from primary owner to delegated owner in order to carry task like bill payment etc. Service roles are also linked to the set of services like broadband, WIFI etc.IAM solution will enable role based privilege management for services exposed to the user. A Mutual trust circle is established between identity provider and services provide to provide Value Added Services.

- ✓ Logical Identity data view can be characterized as Identity profile, Credentials profile, Role/Privilege profile and Service profile. The profile may vary according to on the identity zone model.IAM solution provide extensive amalgamation approach to form single identity model.

- ✓ Entity may have many states, and the journey of traversing through all the states is called as Identity Lifecycle. Identity States may vary based on the zones described above. In a corporate environment, it transits from Day 1 to, Active to, Role assignment to, Role transformation to, Suspension to, revocation to finally Exit. However; in case of social site, not all states are required or applicable.

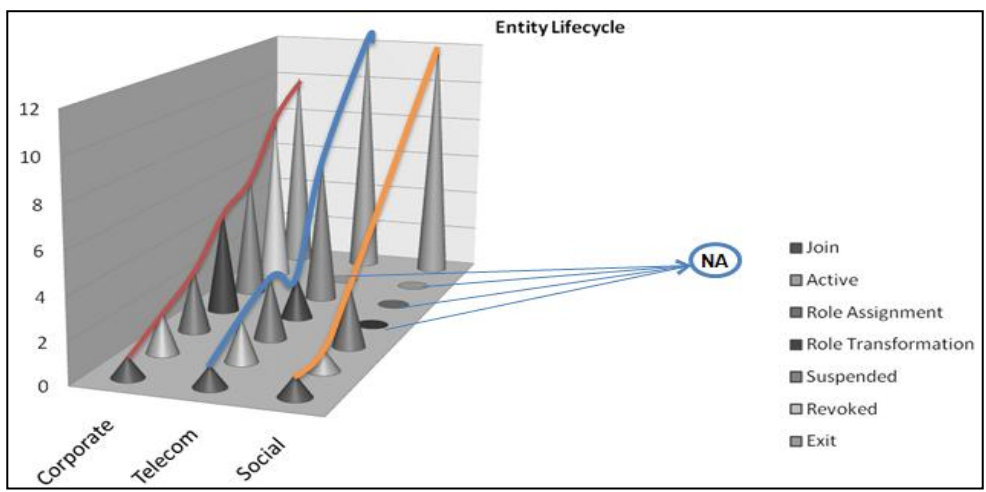

Figure 3: Entity Life Cycle

125

- ✓ Identity may have multiple credentials to authenticate Users for secure privilege management. IAM solution is designed to provide multi-level and multi-factor authentication for segregated identities.
- ✓ Privilege Grant access can be sub divided into Authentication (who AM I) and Authorization (Am I privileged to access the filtered resource). Roles & service association takes major element while defining Role Based Access Control (RBAC).
- ✓ Identity personalization is evolving concept in today's modern world. New Generation Applications are designed to provide value added services based on identity profile.
- ✓ Identities' can be extended across devices' to enable administrative manageability, operational efficiency and security. For example, Physical security, Network device management etc.
- ✓ Minimum sets of information should be disclosed, with the liberty for a subject to modify disclosure agreement in order to keep privacy and legality. A universal identity system must support both "omni-directional" identifiers to be used by public entities and "unidirectional" identifiers to be used by private entities, Thus facilitating discovery while preventing unnecessary release of correlation handles

## 4. PROBLEM WITH DIGITAL IDENTITIES

As discussed above Identity spans many different contexts and purposes: for example, we have multiple individual identity relationships (one with our employer, one with our bank, possibly several with many different parts of the government). There are also role-based identities – a by-product of our current employment, or position. DIZM model also focus on group identities ranging from families through to the companies. To be successful, identity management solutions need to recognize that identity arises from contextual relationships between parties. Customers are expecting a medium to exchange and share information in a *secure, reliable and available (SRA)* mode. Consumers and businesses wanted to simplify the process of logging in to the Websites by creating a single digital identity, with login and password.

Identity of an entity is a dynamic attribute and often changes with regards to systems or environment. The identity Relationship model describes the mapping between Username, Password, Roles and Application.

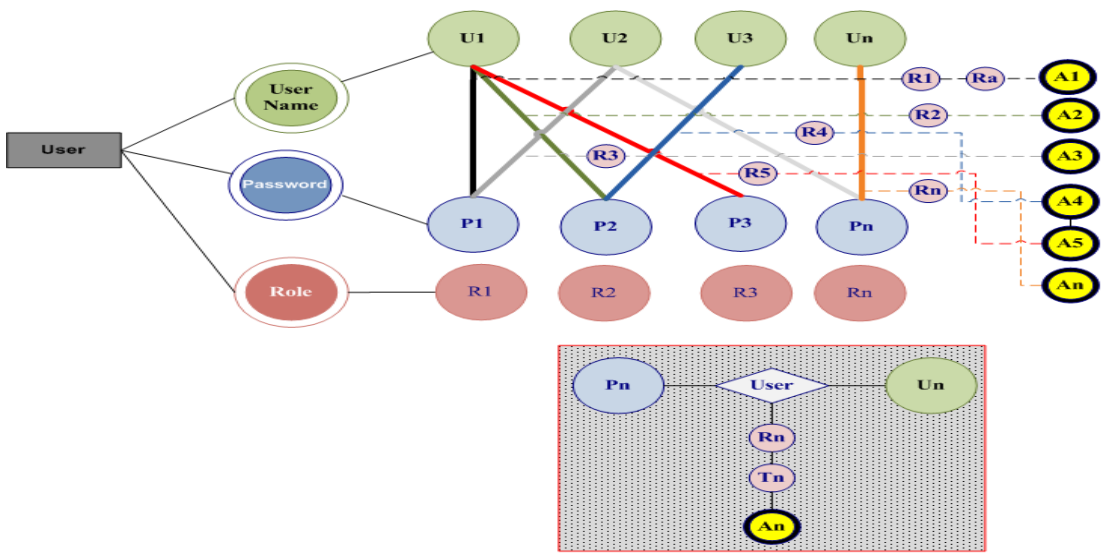

Figure 4: Identity Relationship

## 4.1 Equations

Based on the above Identity Relationship, we are trying to develop a generic user equation to map the DIZM model concept and depict User management complexity

**Terminology Used-**

Let's z represent **Number of Zones** as defined in DIZM, where z>0 and where nz represent Number of Application in nth Zone i.e. Application in $1^{st}$ Zone would be n1, respectively n2….nz, where n>=0

Number of User Id in "z" Zone for "n" Applications=**U**$zn$=Z1 {U1, U2, U3….Un}, where $U_{Zn}$ >= TAzn

Number of Password in "z" Zone for "n" Applications=**P**$zp$=Z1 {P1, P2, P3,….Pn}
Number of Target Application in $z^{th}$ Zone=TAzn

**Analysis**

Set of Unique Userid for n1 application in "$1^{st}$" Zone: **U**$z1n1$ = {($U_{11}$) U ($U_{12}$) U ($U_{13}$)….. U (U$1n1$) }

Set of Unique Password for n1 application in "$1^{st}$" Zone: **Uz1n1** = {($P_{11}$) U ($P_{12}$) U ($P_{13}$)….. U (P$1n1$) }

Total Set of Unique Pair of UserId & Password in "1" Zone=**UPz1n1**= {( $U_{z1n1}$) x (P$z1n1$) }

Total Set of Unique Pair of UserId & Password in "z" Zone=**UP**$zn$**Total**= {( UP$z1n1$) U (UP$z2n2$) ………… U (UP$zn$)}

**Credential Equation for User = {UPznTotal}**

Only UserId and Password is considered as a authentication parameter (Exclusion biometric, certificate etc) for the Credential analysis

Activity performed by the user is consider as a Task (t), which can be mapped as $Tasks \in Roles$, where Task >0 for any set of application where user has access.
We are assuming, different Target application within same zone of DIZM, may have similar task associated with the Roles.

Set of Unique Roles "m" (where m=no of Roles) in "$1^{st}$" Zone for n1 Applications: **R**$_{z1r1}$ = {($R_{1a}$) U (R $_{1b}$) U (R $_{1c}$)….. U (R $_{1m}$) }

Total Set of Unique Roles in "z" Zone=**R**$_{zr}$= {( R$_{z1r1}$) U (R $_{z2r2}$) ………… U (R $_{znrm}$)}

**Role Equation for User={(Rzr)}**

The above model emphasizes on multiple identities managed across various zones and hence a user may needs to remember huge number of credentials to securely access. YAUP (Yet another

Username/Password) has lead to storing credentials in a piece of paper or a file. This is a major concerns or security risk which hacker are exploiting

## 5. Technology Solution

Technology Evolution is defined in order to established characteristics of identity from Stone Age to as in state. The revolution has crossed various transition phases like Reference/Paper based, Evolution of Digital Identity, Multiple Identity silos due to technology expansion, User Centric conversion & User Centric conversion with Trust Assurance model.

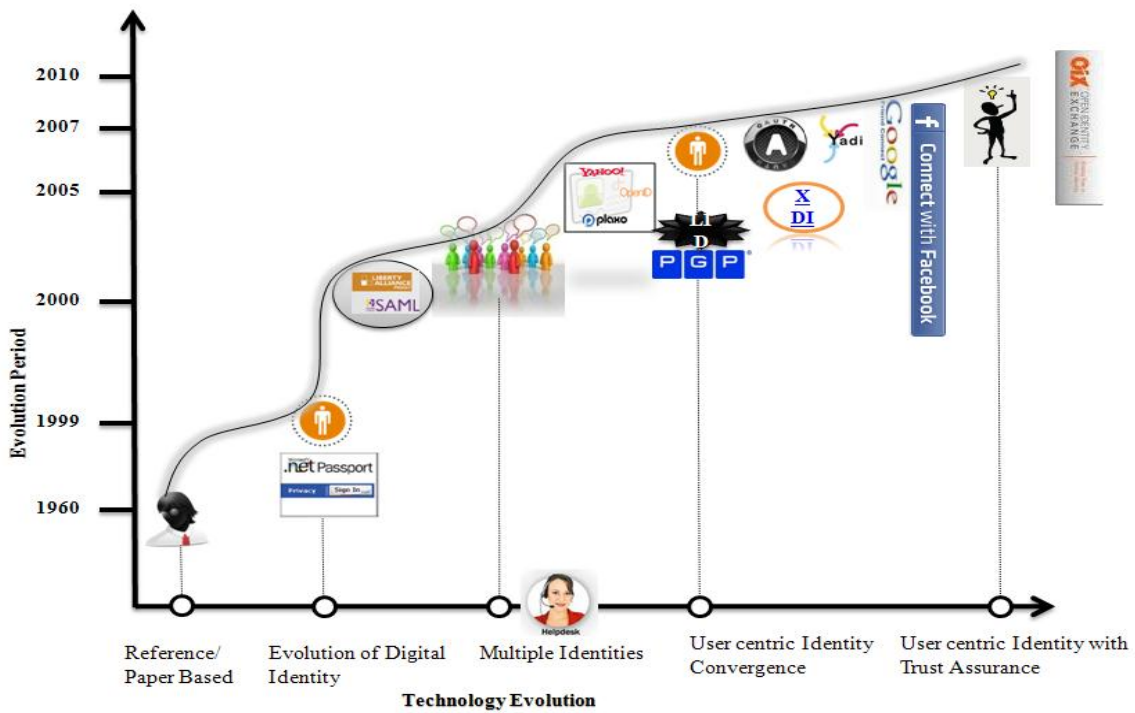

Figure 5: Evolution of Technology

The technology evolution will be discussed using technology Quadrant approach. Four planes are considered to concentrate on the problems associated with decentralized identity and data over the meshed network. Also, it highlights the transformation path with respect to Service Transformation, Authentication & Technology conversion in technology Quadrant.

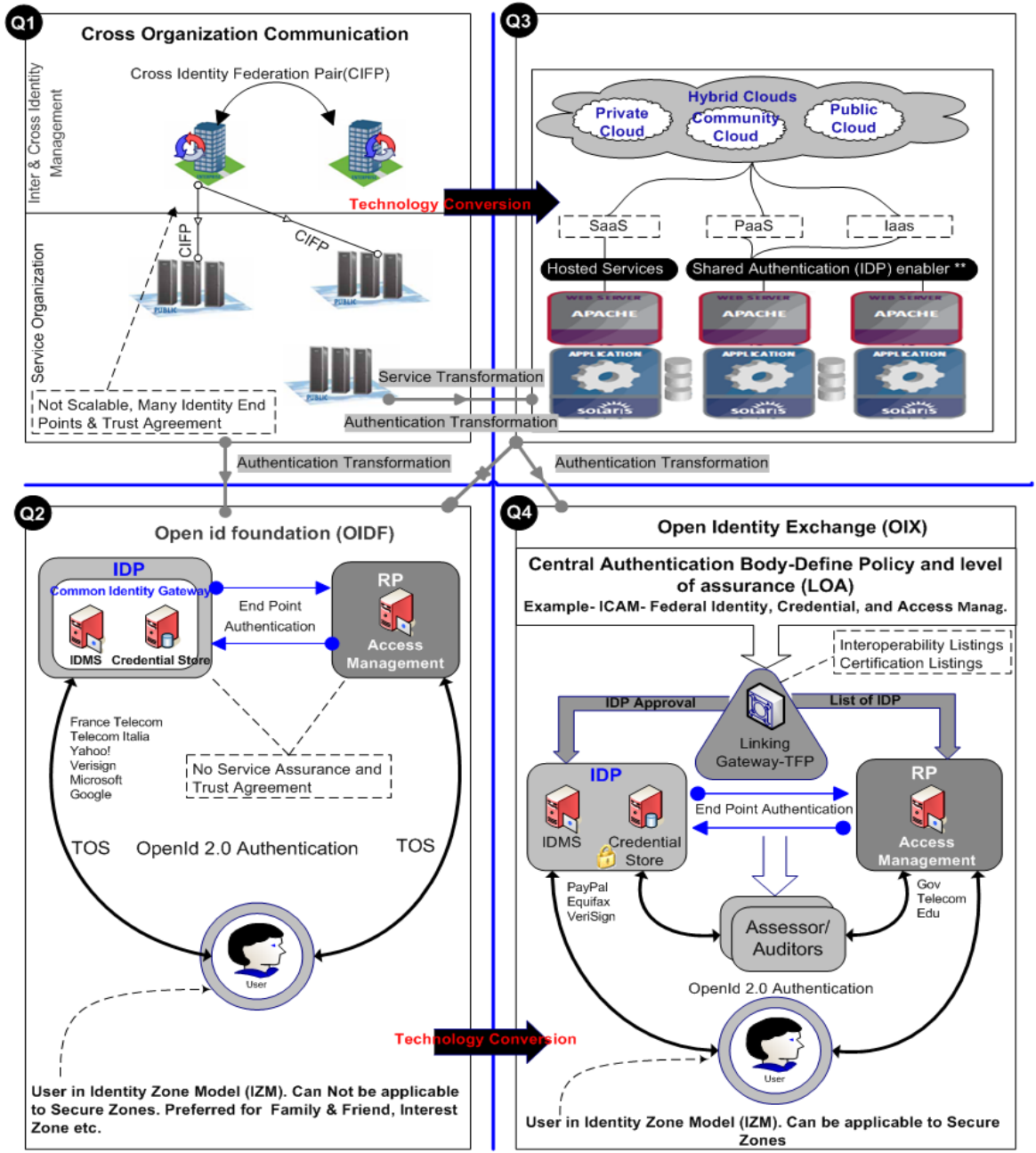

Figure 6: Technology Quadrant

1. **Q1 Represent**- Inter & Cross identity management: Cross Identity Federation Pair (CIFP) & oAuth enabled protocol.
2. **Q2 Represent**- OpenId & oAuth mechanisms to symbolize decentralized internet identity needs.
3. **Q3 Represent**- Cloud computing. It can be extended to enable IDP, RP & oAuth Service provider infrastructure.
4. **Q4 Represent**- Open Identity Exchange. It provides Assurance and Trust framework to support Single identity model for secure zone.

Based on the above Technology Quadrant, following solutions or approach can be defined-

## 5.1 How to share identity in Decentralized Internet ecosystem

**Cross Identity Federation Pair (CIFP):** Information can be securely exchanged between different entities using a one to one trust association. Participating organizations is denoted as IDP and IDC is Cross identity federation pair (CIFP) to achieve seamless SSO .Service organizations are actively participating in CIFP model, using a common Identity Gateway. Typical example would be large telecom to provide NGN service offerings. Common identity can be used, however it lags when compared with a bigger model like DIZM and will not be a scalable solution to consider. In this zone **COTS** based Identity & Access management solutions (Oracle, CA, IBM etc) are primarily used and focused upon. Various authentication protocols like SAML, Liberty etc are consider to securely transform identity information across **CIFP.**

## 5.2 How to enable Data portability along with Identity consolidation?

There are two approaches to access data stored across various sites. Authenticate user at all the site (with Single User Id & Password) or Authenticate user at one site and use the security token in remaining site.oAuth is based on API based authorization mechanism which works in backend to access the segregated data without sharing password.

**Data portability via OAuth:** oAuth protocol lies under LWI protocols. This is primarily used to share data segregated across various sites in internet ecosystem. Classic example would be google & Twitter that provide OAuth Integration mechanism. Following options are supported to securely share data without Federation based infrastructure. oAuth provides a backend channel to retrieve the segregated data using application API's. Facebook Connect is a new service that allows users to login other websites like CitySearch, CNN and even share their activities from these third-party sites with their Facebook friends. Google has a similar service named Friend Connect, launched earlier. Friend Connect does the same thing as Facebook Connect, but only for users of social networks viz Orkut and Plaxo. oAuth based model can be used to integrate with Online Web photo album sites with Photo Printing sites. User doesn't need to share his credentials with Photo printing site to avoid privacy issue.

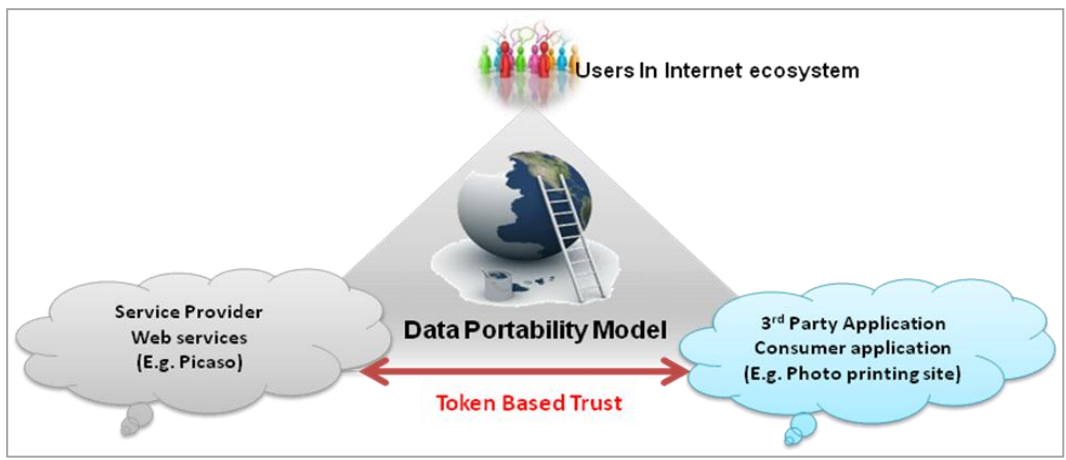

Figure 7: Data Portability

The objective to describe oAuth is to map identity profile & identity data over the internet. A dedicated white paper would be required to address the acceptance of oAuth Framework.

### 5.3 How to share identity in Decentralized Internet ecosystem with User Centric Model-

This is a bigger model as compared to CIFP, with many IDP and RP interacting, focusing on user desire to share information.

*"It's a user-centric identity system that provides more control to users' of how their information is shared".*

**OpenId**-OpenID is a classic example to address the user centric model; however the issue is to enable trust between Identity providers and relying parties. OpenID has many benefits like User ease along with integration interoperability with vast Access management products (e.g. Opensso can be extended with OpenID extensions called as Provider Authentication Policy Extension (PAPE) ).Sun Microsystems has joined the league of identity provider for employee with mandatory Registration phase. Critical component to establish OpenID based model is by considering Registration framework which will act as profiler. User centric model provides an enhanced capability to manage credentials centrally but hence it can be central point of security breach. OpenID based ecosystem can leverage the enhanced security by enabling multifactor and or PKI based solution.

More than 9 million Web sites currently accept OpenID for around 1 billion users

In OpenID enabled architecture, the endpoints are typically called as Relying Party (RP) and the Identity Provider (IdP). Figure [6] represent the Identity provide, which manages identity profiling and acts as an Individual central authentication body (ICAB).RP redirects the user to appropriate IDP based on OpenID URL(**Username.IDP_NAME.COM**) OpenID 2.0 supports the following features: single sign-on, session reset, attribute exchange, pseudonymous identifiers, and authentication policy. User has a liberty to navigate either from Identity provider (IdP) end points to Relying party (RP) or vice versa. From IdP, user can manage the session, central Interface management, comprehensive Auditing features. OpenID authentication 2.0 can be extended to delegate authorization capability to RP based on profile attribute in assertion profile. User has the liberty to specify if the information to be disclosed with the RP as per disclosure agreement. OpenID based system maintain vast session history for user to track the activities performed.

Legal, operational, compliance and Business requirements are hard to enforce without any mutual legal agreement. Businesses are finding it difficult without any policy enabled trust framework in place.

### 5.3.1 Technical Limitations of OpenID system when discussed around bigger model like DIZM

1. Secure Communication is a major concern, since OpenId system works on Authentication based on redirection approach. This could lead associated vulnerabilities like Phishing attack, MITM Attack, Impersonation etc. Also, underlying system vulnerabilities are hard to discover from RP.
2. How common registration at IDP meets the needs of RP specific profile information. A global registration process should be constructed which will validate the user during User at Registration Time.
3. Who will manage Password Policy for huge number of applications, since we are talking about Multiple IDP based system? How application security Policies are mapped to IDP Framework.
4. How existing accounts of RP will be mapped to the OpenID account. RP needs to maintain relationship between exiting accounts, Open Id accounts, Roles and service profiles.
5. No policy enabled framework, which makes difficult to venture into secure web sites like government. Also, there are no standard to manage Entity lifecycle as defined in Figure [4]

6. User experience & Privacy Concerns: OpenID uses a HTTP URL to authenticate user, which is not in line with traditional username and password mechanism. The acceptance for Non technical users is negligible. User has to remember 2 different login credentials- one for email & other for remaining web sites. There are many suggestions like using an Email as a common credential. Using Email address as OpenID URL (username@address.com rather than http://OpenID.address.com/username) is an option but has many privacy related issues.
7. IDP should discourage Re-allocation of Your OpenID to different user if not been used for long.

### 5.3.2 Security Countermeasure

This section list few security related countermeasure which can be consider
1. To have either one time password or Token based authentication mechanism to mitigate Security Risk like Man-In-The-Middle Attack
2. Use of SSL enabled protocol to limit Phishing attack. However this will not entirely eliminate the occurrence, since its likelihood depends to User awareness.

## 5.4 How to enable outsourced Service & Authentication transformation-

**Cloud Computing (CC):** Cloud computing is an emerging trend, to mitigate the needs of huge infrastructure for hosting service offerings.

*Simple definition of Cloud: - "On demand usage of compute and storage over virtualized environment"*

As part of User centric approach organizations need to invest massive amount in order to enable IDP. CC makes use of SaaS, to effectively use shared environment for enabling IDP infra.CC has been going through many issues, arising from compliance requirement to ensuring trusted mechanism for customer. Cloud can be Classification as - Private, Community-based, Public and hybrids. It can be further broken down into various outsourced solution like SaaS, PaaS or IaaS. IDP, RP and service provider can be easily hosted in CC enabled infrastructure. Cloud offers 4 deployment models: Public, Private, Hybrid, and Community.

Figure [6] represent Technology Conversion, Service Conversion and Authentication transformation from various Technology quadrants to Cloud Computing (CC).Services and technologies are mounted and leveraged upon by outsourcing model.

CC has many areas to look upon for universal acceptance around the globe. We are not going to address in detail as part of current discussion.

### 5.4.1 Few Concerns with cloud enablement

1. Governance Issues pertaining to service assurance

   - SLAs assurance and measurement- Recovering true cost of a breach: penalties vs risk transference

   - Poor business continuity planning

   - Cloud providers and customers need defined collaboration for incident response.

2. A uniform & comprehensive compliance standard (SAS 70 II, ISO 2700X) required for assessing wide scoping of cloud.

   - No audit standards specific to the 'cloud'

   - Most cloud providers don't even have a SAS-70.

   - Maintain a right to audit on demand

3. Land of Law

   - Patriot Act Problem -Data Centers in countries with unfriendly laws

   - Cross-border data transfers

   - Data Storage mechanism

4. Evolving Threats landscape

   - Unprotected APIs / Insecure Service Oriented Architecture, Hypervisor Attacks,L1/L2 Attacks (Cache Scraping), Trojaned AMI Images, VMDK / VHD Repurposing, Key Scraping, Infrastructure DDoS, SRF,XSS,SQL Injection, Data leakage, Poor account provisioning, Cloud provider insider abuse, Click Fraud etc.

## 5.5 How to share an identity in Decentralized Internet ecosystem with Service Assurance and Trust Assurance

### Open Identity Exchange (OIX)

The concept of a single online trusted identity has started with OpenID, but was not very popular due to its limitation around assurance and trust. As part of US initiative a new nonprofit group called the Open Identity Exchange (OIX) is providing a trusted framework for the exchange of online identity credentials on the public Internet and in private data communications. Quadrant 4 of Figure [7], provide interaction between various parties along with applicability of the solution with respect to DIZM. This is an extension of OpenId framework, where more emphasis is given on Security along with process enabled framework (Interoperability listings etc).

This approach enables a scalable model for extending identity assurance & Trust across business needs. This model can be extended to support other industry specific needs arising due to converging nature of technology.

> **This model can be used as a central authentication body and can  work in conjunction with Number portability program in Telecom sector**

The National Institute of Health is the first government agency to accept OIX logins. Objective of OIX based model is to define compressive framework to address national needs like standards, regulation and laws, minimum risk, interoperability and to provide user with a consistent context or user experience at government site. The Open Identity Trust Framework (OITF) Model, does not require RP to have credential management services, since those features are "outsourced" to the IdP with a defined Operational, Technical & legal requirement to enable secure trustable digital information.

In Open Identity Trust Framework (OITF) Model, IDP's and RP's interact via Policy Interoperability & Technical Interoperability define as part of framework. This will overcome the limitation of Openid based model described in section [4.3.1].

We are here describing the same model as used in US to support global identity needs. The solution is divided in 4 parties actively participating –

1. **Central Authentication Body (CAB)-**This committee can be named as Policy Makers and are members who define a LOA and LOT guidelines. These policies will be well accepted across all the verticals like Government, Telcom, education etc. However, based on the vertical, different policy maker committee may be defined to example- for government web site, Government Policy maker as CAB, for telecom-Industry association etc. Now the question is how to address individual national needs, if this is going to be central committee. Are we going to have separate Policy maker –The answer is may be In Future, individual members of respective nation could also become a member of CAB or may run individual CAB's(nationalized). CAB can be defined as the heart of OIX model, since they define operation and legal aspect for communication between vast number of IDP's & RP.
2. **OITF Providers**- (OITF Providers) transform the requirements of policymakers into their own blueprint for a trust framework that they then proceed to build. As OITF Providers do so, they need to attract parties by explaining how their requirements support the interests of all.
3. **Assessors** evaluate IDP and RP to certify that they are capable of following the OITF Provider's blueprint.
4. **Auditors** will evaluate the practices carried between the parties based on what was agreed for the OITF.
5. **Legal Disputes** may be catered via dispute resolution services.

## 6. CONCLUSION:-

Technology revamp has driven from user ease, manageability and Cost efficiency. Identity is transient from many phases like, single digital identity, multiple segregated identities and again a single consolidated identity model via OpenID & OIX. Internet is a meshed network with many identities segregated over DIZM model. This has lead to Criminalization of the Internet via loose coupling and hence defects in the systems. Many Non-Profitable organizations have come up with an idea to develop a common framework to share identity and data in a secure and reliable mode. OpenID is one of the solutions used for more than 9 million Web sites and 1 billion users' base. Many organizations are actively showing interest in order to enable consolidated identity framework, however Legal, operational, compliance and Business requirements are hard to enforce without any mutual Legal agreement. Businesses are finding this state difficult to achieve without any policy enabled trust framework in place. Trust & assurance have given more importance in developing single identity mode like OIX. This is still not mature and it's in an initial phase of definition and there are still few questions unanswered and need more acceptances globally like

1. Universal Registration Process-It would be stringent to accommodate Universal registration process applicable to all zones in DIZM. Individual RP applications may require separate registration processes for Application authorization.

2. Universal acceptance: This may be implemented across major industry segments like Government, Telecom etc.

3. Globalization-Legal constraints (Laws of Land) in sharing identity information for global applicability across nations.
4. Balance between Single Point of Failure & Security Risk. Probability of such an occurrence is negligible but cannot be avoided.
5. Single Identity may not be applicable to the entire zones defined in DIZM model. Separate identities may be required for individual zone to cater Information data classification.

**References**

[1] Open Identity Exchange Framework(OIEF) Model - http://openidentityexchange.org

[2] List of Openid Providers on the Internet: http://en.wikipedia.org/wiki/List_of_OpenID_providers

[3] Don Thibeau, "A Market Solution to Online Identity Trust" .Available from www.itu.int/dms_pub/itu-t/oth/06/35/T063500000200301PPTE.ppt

[4] OpenID Foundation (OIDF), http://openid.net

[5] OpenID. "Microsoft and Google announce OpenID support". 30 October 2008. See http://openid.net/2008/10/

[6] D. Recordon B. Fitzpatrick. "Open ID Authentication 1.1". May 2006. Available from http://openid.net/specs/openid-authentication-1_1.html

[7] Project Director-Peter Burnhill. "JISC Final Report- OpenID Study". 30 November 2008. Available from http://www.immagic.com/eLibrary/ARCHIVES/GENERAL/JISC_UK/J081203F.pdf

[8] Waleed A Alrodhan. "Privacy and Practicality of Identity Management Systems". 17 November 2010. Technical Report- RHUL-MA-2010-14. Available from www.ma.rhul.ac.uk/static/techrep/2010/RHUL-MA-2010-14.pdf

[9] Mehrdad Naderi1, Jawed Siddiqi1, Babak Akhgar1, Wolfgang Orth2,Norbert Meyer3, Miika Tuisku4, and Gregor Pipan5. "Towards a Framework for Federated Global Identity Management". Vol.7, No.1, PP.88-99, July 2008.

[10] Wayne Jansen & Timothy Grance, "Guidelines on Security and Privacy in Public Cloud Computing". NIST Draft Special Publication 800-144. Available from http://csrc.nist.gov/publications/drafts/800-144/Draft-SP-800-144_cloud-computing.pdf

[11] Sponsored by CA, "Security of Cloud Computing Users". Available from http://www.ca.com/files/industryresearch/security-cloud-computing-users_235659.pdf

[12] Beverly Freeman," OpenID User Experience Research". July 2008.Avaiable from http://developer.yahoo.com/openid/openid-research-jul08.pdf

[13] Microsoft, "The Laws of Identity", http://msdn.microsoft.com/en-us/library/ms996456.aspx

[14] Radovan Semančík, "Enterprise Digital Identity Architecture Roadmap". April 2005, Available from http://storm.alert.sk/work/papers/nlight/enterprise-digital-identity-architecture-roadmap-v1-2.pdf

[15] Jolie O'Dell. "How Much Does Identity Theft Cost"? January 29, 2011. Available from http://mashable.com/2011/01/29/identity-theft-infographic/

[16] Phillip J. Windley," Digital ID and eGovernment" , "Understanding Digital Identity Management" , http://www.windley.com/docs/Digital%20ID%20and%20eGovernment.pdf, http://www.windley.com/docs/Digital%20Identity.pdf

[17] Rob Richards, "Authorization with OAuth", October 22, 2009. Available from http://cdatazone.org/talks/zendcon_2009/OAuth.pdf

[18] Alvaro J. Gutierrez, "Towards Better Digital Identity Management". CPSC 457b - Spring 2006

[19] Sean P. Aune, "25+ Ways to Manage Your Online Identity". September 10, 2007. Available from http://mashable.com/2007/09/10/online-identity/

[20] Sean P. Aune. "Kim Cameron, "The Laws Of Identity", http://www.identityblog.com/?p=352". September 10, 2007.

[21] Rachel Kim, Analyst, "2009 Identity Fraud Survey Report: Consumer Version". February 2009. Available from http://www.search.org/files/pdf/IdentityFraudSurveyConsumerReport.pdf

[22] Privacy Rights Clearinghouse, http://www.privacyrights.org/fs/fs17g-CrimIdTheft.htm

[23] RNCOS. Identity and Access Management Market Forecast to 2012 (research report), February 2009. http://www.rncos.com/Report/IM181.htm.

[24] Janrain. "Consumer Perceptions of Online Registration and Social Sign-In". US Consumer Market Research. http://www.janrain.com/

**Authors**

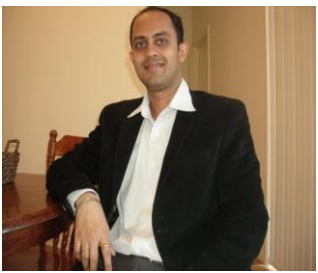

**Mohit Kohli** received the B.E., from Mumbai Univ. India in 2001.He Started his carrier in Wide area network Technology solutions. He headed various complex Security projects in service organizations like TechMahindra Ltd. and Siemens Information system Limited. Currently, he is heading the Security Practice of Rolta Technology. Also, he is certified professional - Advance Sun certification for Identity & Access Management, Certified Ethical Hacker and Countermeasures (CEH-V 4.0: EC Council), BS7799 Lead Auditor, Systems Security Engineering (SSE-CMM), Checkpoint Certified Security Administrator (CCSA) and Cisco Certified Network Associate (CCNA).His research interest includes Security- Identity and Access Management solutions and Risk Management etc.